\documentclass[twocolumn]{aa}
\usepackage{graphicx}
\usepackage{txfonts}
\begin{document}
\title{Speckle observations of binary stars with a 0.5 m telescope}
\author{A. Rutkowski\inst{1} \& W. Waniak \inst{2}}
\offprints{rudy@camk.edu.pl}
\institute{ N. Copernicus Astronomical Center, Bartycka 18, 00-716 Warsaw,
Poland\\
\and
Astronomical Observatory of Jagiellonian University, Orla 171, 30-244
Cracow, Poland\\ }
\abstract{We present 36 observations of 17 visual binaries of moderate
 separation (range from 0.15$''$ to 0.79$''$) made with the 50 cm Cassegrain
 telescope of the Jagiellonian University in Cracow. The speckle interferometry
 technique was combined with modest optical hardware and a standard photometric
 CCD camera. We used broad-band V,R,I filters without a Risley prism to reduce
 differential colour refraction, so we performed model analysis to investigate
 the influence of this effect on the results of measurements. For binary
 components of spectral type O-F, the difference of three spectral classes
 between
 them should bias their relative positions by no more than a couple of tens of
 milliarcseconds (mas) for moderate zenith distances. The statistical analysis
 of our results confirmed this conclusion. A cross-spectrum approach was
 applied to resolve the quadrant ambiguity. Our separations have RMS deviations
 of 0.012$''$ and our position angles have RMS deviations of 1.8$^{\circ}$. Relative
 photometry in V, R and I filters appeared to be the less accurately determined
 parameter. We discuss our errors in detail and compare them to other speckle
 data. This comparison clearly shows the high value of our measurements.
 We also present an example of the enhancement of image resolution
 for an extended object of angular size greater than the atmospheric coherence
 patch using speckle interferometry techniques.

\keywords{Binaries: visual - Techniques: interferometric - Techniques: high angular resolution - Astrometry}
}
\maketitle

\section{Introduction}
Although during the last decade adaptive optics has pushed the once
prominent technique of speckle interferometry (Labeyrie 1970) into the
background,
this method still provides a simple, inexpensive tool for determination of the
orbital and photometric parameters of visual binary stars in the course of
surveys made with small and moderate sized telescopes. Many such projects
have been
completed in the last years (e.g. Mason et al. \cite{mason-a}, \cite{mason-b},
Prieur et al. \cite{prieur-b},
Scardia et al. \cite{scardia}). Substantial technical and programming improvements,
as well as the widespread use of CCD cameras among small observatories and
amateurs, make it possible to perform both speckle observations of binary stars
(with quite good precision) and also diffraction-limited speckle imaging of
compact extended objects.

In this paper, we show how a standard photometric CCD camera usually
used for broad-band photometry of variable stars and comets, can be
inexpensively modified into an uncomplicated "specklegraph", which can supply
measurements of binaries hardly inferior to the results of
observations with a 1-m class telescope equipped with classical speckle cameras
which include reducers of differential colour refraction and/or narrow-band
filters. For the trial run we selected a number of binary stars with
separations ranged from one half of the diffraction limit of the telescope for
visible light up to 1$''$ and brighter than about 8 mag in V. Some of them have
relatively precise orbital parameters (e.g. WDS15232+3017), some have extremely
inaccurate orbits (e.g. WDS11137+2008), whereas others are still waiting for
such a determination.

In the following paragraphs we describe our equipment, observational
procedure and the reduction method of speckle images. Before presenting
our results we  discuss the problem of the influence of
differential colour refraction. Conclusions reached from model
computations are confronted with the observations. We present orbital O-C
analyses for pairs with well known orbits. We also show an example of how the
image resolution degraded by seeing can be improved for extended objects with
angular dimensions greater than the radius of the atmospheric coherence patch.
\section{Equipment}

The observations were performed using a Cassegrain telescope (0.5 m diameter
and 6.7 m  focal length), with a PHOTOMETRICS S300 CCD camera with a SI003B
thinned, back-illuminated chip manufactured by SITe. It has 1024x1024 square
pixels of the side 24 $\mu$m and a 16 bit A/D converter. The camera works
in the Multi Pinned Phase (MPP) mode with a thermoelectric+liquid cooling
system, which maintains operating temperature close to $-50^{\circ}C$.

We observed with the set of standard V,R,I filters specially designed for
CCD (Bessell \cite{bessell} ) manufactured by CUSTOM SCIENTIFIC. U and B filters were
not used for two reasons. The first was the diminishing of the useful
signal due to relatively low quantum efficiency of the CCD in this range
and atmospheric
extinction. The second was the possibly high bias of the measured
positions of the binary components by differential colour
refraction. Model analysis of this effect confirmed our approach.

The typical angular scale of the chip, equal to 0.736 arcsec/pixel, is
insufficient for proper sampling of a specklegram, so we used a projector
giving appropriate enlargement. This has the form of a duralumin tube of length
 150 mm and inner diameter 40 mm, attached to the back side of
the filter wheel chassis. The CCD camera is mounted to the back flange
of this tube. The optimised connecting system assures straightforward and quick
(up to a minute) installation of the projector. Great attention was paid to
the removing rotational backlash. Inside the tube, typical
microscope objectives (10x or 20x) can be mounted interchangeably.

The angular scale of our "speckle camera" for the 10x objective was obtained
from astrometry of $\alpha$¡ Gem (Castor, WDS07346+3153=ADS 6175) and relative
positions
from Hipparcos data. Separation of the components (close to 4$''$) exceeding
separations of our target pairs and close proximity of spectral classes
(A1V, A2V) minimising differential colour refraction make this object highly
suitable for astrometric calibration. In the case of the 20x objective we used
combined results from speckle interferometry of Castor and
WDS15232+3017=ADS9617. The latter star has precisely obtained orbital parameters
(grad 1), and is recommended as an astrometric calibrator by Hartkopf \& Mason
(\cite{hartkopf-a}).
The scales of our frames were equal to 0.0824$\pm$0.0002 and 0.04760$\pm$0.0007
arcsec/pix for the 10x and 20x objectives, respectively. The precision of
determination of position angle was close to 0.1$^{\circ}$.
This error was the result of a
small rotational backlash in the focusing head to which the whole assembly of
the filter wheel, the projector and the camera was mounted. We did not notice
any serious image deformations which could be given by the projector.

As our CCD camera is typically dedicated to precise photometry it has low
Read Out Noise (RON)
but relatively long read-out time of 13 sec per frame. Taking into account the
precision of our autoguider system and the relatively bad seeing (frequently
exceeding 3$''$) we decided to use a sub-frame of 256x256 pixels,
where the readout
time was close to 2 sec. Comparing the readout frequencies of typical speckle
cameras (tens of cycles per second) with  our system it is obvious that our
observations have low effeciency. On the other hand they ensure full statistical
independence of consecutive frames for larger and longer-lasting wave-front
disturbances. An attempt to use the promising drift scanning mode
(Fors at al. \cite{fors-a}) enhanced our speed only by a factor of two while
producing prominent streaking between speckle images. In the end we used the
common mode of data transfer.

\begin{figure}
\resizebox{\hsize}{!}{\includegraphics[bb = 55 175 520 630,clip]{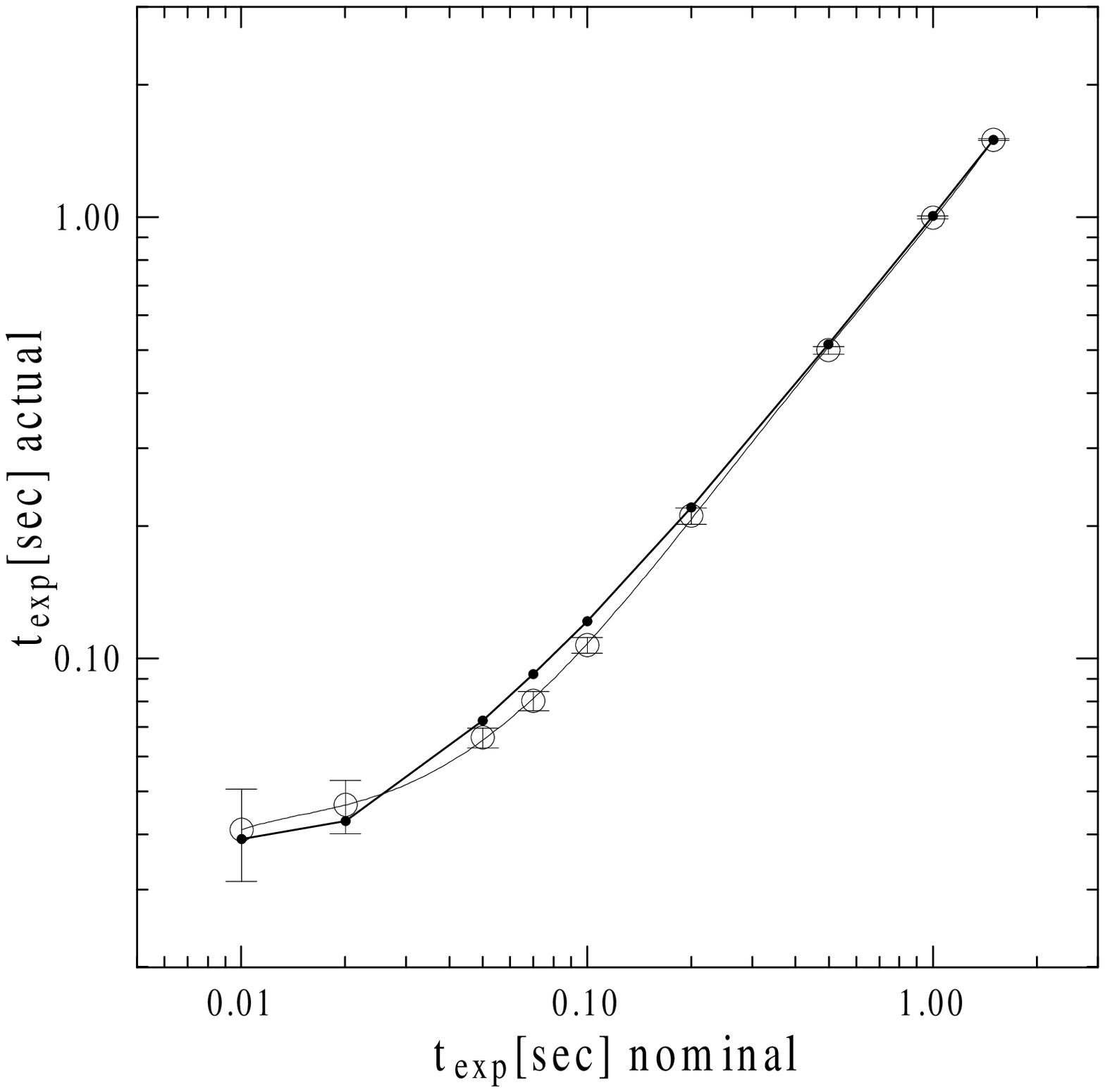}}
\caption{Dependence between actual (achieved) and nominal (set by the observer)
exposure times for the iris shutter of the CCD camera. Thin solid line: 5th
order polynomial fit to the data. Thick solid line with dots: simple mechanical
model of the shutter.}
\label{Fig1}
\end{figure}

Of crucial importance for the proper realisation of speckle interferometry
is the
fulfilment of the condition of frozen turbulence. This means using extremely
short exposure times (of the order of milliseconds).
Our electronically controlled
iris shutter has a nominal exposure time of up to 1 msec, but its dead time
is of the order of 50 msec. Therefore, we were forced to determine actual
achieved
exposure times and investigate their repeatability. We obtained images of a
bright star using nominal exposure from 2 sec (where the dead-time effect is
invisible) up to 10 msec. The results are presented in Fig.~\ref{Fig1}.
The conclusion is
that the shortest achievable integration time is close to 40 msec, quite
comparable with the dead-time of the shutter. Analysis of the
dependence observed in this figure using a simple model which included the
mechanical inertia of the shutter showed that the magnetic hysteresis of
the electromagnet should be taken into account.
\section{Observations and reductions}
Speckle observations were performed during 13 nights from March
to June 2003. Almost all binary stars satisfying the conditions described in
the first paragraph and visible at the appropriate conditions during this period
were studied. We tried to observe as large fraction of our sample as possible
with different PSF standards, filters and exposure times to get a feeling of
how the obtained parameters depend on them.

A typical observing run consists of a series of 100 dark frames, 300 frames of a
PSF standard (one half obtained before, and one half after the observation of
the binary) and 300 frames of the object of interest. We realised two to three
such series for each star per night. The duration of a single run was shorter
than about half an hour, which ensured limited variation of the seeing
conditions.

Flat frames were obtained on the twilight sky using exposure times the same as
were used for observing the binaries. This was extremely important for taking
into account the non uniform illumination of the field of view given by
the iris shutter for extremely short exposures. PSF standards were chosen as
close on the sky to the binary stars as possible, considering the zenith
distances and the spectral classes. Nominal exposure times ranged from 10 to
100 msec, which means that achieved integration times were from 40 to 107 msec.
Generally, we correlated them with the brightness of the binary and PSF standard
and the separation of the components. Shorter exposures imply better
contrast of the speckle structure but a lower signal level. Anticipating
discussion of the results, we can ascertain that using different integration
times we obtained astrometry of the same class of precision.

We started the data reduction with subtraction of the mean dark frame,
which for such short expositions mainly carries information about the bias level.
After flat-fielding of the object frames, we shifted the images of the speckle
patterns to the central position on the frame, to facilitate later
 steps of the data processing. Substantial chaotic transitions of the
pattern were produced both by wave front tilts and autoguiding process, which
had an overall precision of 1$''$². Next we conducted apodization of the frames
with a smooth circular mask. The unmasked region had a diameter large enough to
encircle the whole visible speckle pattern while small enough to cut off as
much of the noisy surroundings as possible.

After computing the mean Fourier power spectra of the images of both binary stars
and PSF standards, we subtracted from them the mean power spectrum of the dark
frames apodized by the appropriate mask as well as photon noise bias
(Dainty \& Greenaway \cite{dainty}) estimated by the mean level of the spectra
beyond the diffraction limit. At this stage of the data reduction
process we compared the spectra of the PSF specklegrams observed before and after
the binary. In the case of full agreement between them we combined both results
into one set, which was reanalysed to give the final power spectrum
of the standard.
The spectrum of the binary was then divided by the PSF spectrum treated as a
Wiener filter (Keller \& Johannesson \cite{keller}). The distance between binary
components, the position angle and the relative brightness were obtained from
LSQ fitting of a sinusoid to the resultant power spectrum. We masked the
central part, where the seeing peak dominates, and the outside regions,
where the S/N
ratio is too low a the cut-off of spatial frequencies greater than the
diffraction limit takes place. No other weighting of the input data was
applied. After close inspection of the images we discarded the classical
$\chi^{2}$  approach (Horch et al. 1997) as the photon and read-out noise were
not only source of the variance seen in the power spectra.
Quasi-systematic horizontal noisy structures played a quite serious role
in our dark frames. They
produced weak vertical structures in the images of the power spectrum,
structures which were only partially removed by the subtraction procedure.

\begin{figure}
\resizebox{\hsize}{!}{\includegraphics[bb = 150 675 465 775,clip] {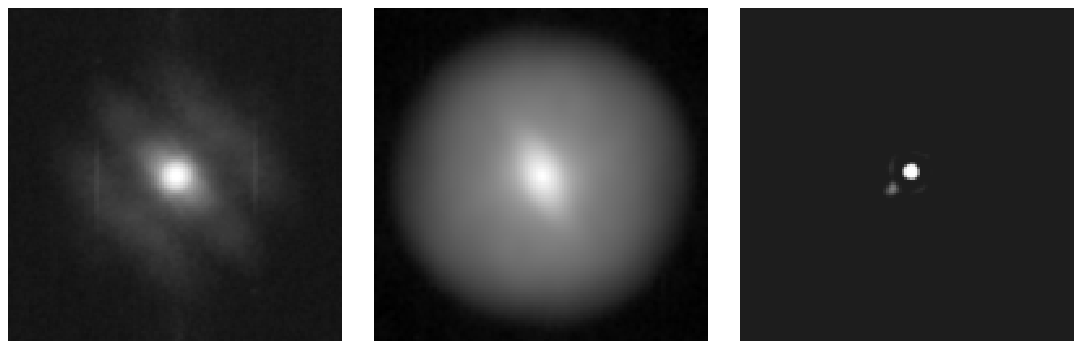}}
\caption{Example of speckle imaging for WDS 04100+8042=ADS 2963.
Left panel: raw spectrum of the binary, middle panel: spectrum of the PSF
standard, right panel: image reconstructed by cross-spectrum analysis.
Separation between components is 0.78 arcse. North is up and East is to the
left.}
\label{Fig2}
\end{figure}

To resolve the quadrant ambiguity, we used the cross-spectrum approach by Knox
\& Thompson (1974) (see also Christou \cite{christou}) analysing 8 sub-plains computed
for 8 shifts from a given pixel towards all it's neighbour. The resultant
frame bearing phase information was obtained by an iterative LSQ approach. In the
case when the distance between binary components was greater than the
diffraction limit, we were able to obtain two separate peaks. For closer
components the asymmetry of the elongated profile of three overlapped peaks
(one central and two satellites) made it possible to resolve the quadrant
ambiguity. An exemple of the result of our speckle imaging procedure for the binary
star WDS 04100+8042=ADS 2963 is presented in Fig.~\ref{Fig2}.
\section{Theoretical differential colour refraction}
Prior to presenting the results of our speckle interferometry, we must
discuss in detail the effect of differential colour refraction (DCR). As our observations
were performed without a Risley prism corrector and with broad-band V,R,I filters
we must pay great attention to this phenomenon.

The dependence of colour refraction on the atmospheric
temperature, pressure and relative humidity as well as spectral classes of stars
(taking into account broad-band photometry systems) has been investigated both
theoretically and observationally (e.g. Stone \cite{stone-a}, \cite{stone-b}, Gubler \& Tytler \cite{gubler},
Malyuto \& Meinel \cite{malyuto}, Langhans et al. \cite{langhans}). Computing the DCR effect we used
formulae for atmospheric refraction taken from Stone (\cite{stone-a}). The only
modification was the replacement of the spectral energy distribution of the star
by the spectral distribution of the number of photons, which is suitable for
quantum sensitive detectors such as CCDs. We used the dependence of the quantum
efficiency upon wavelength and the transmittance curves of our filters as given by
the manufacturers. The transmission of the atmosphere was reconstructed from
the mean extinction coefficients in the U,B,V,R,I bands (Siwak, private
communication), taking into account the absorption by
ozone and both Rayleigh and aerosol scattering. We assumed the transmittance of
interstellar matter characteristic for the colour excess E(B-V)=0.3, although
we examined the DCR effect for other values. We used the spectral energy
distributions for appropriate spectral classes taken from the BaSeL Stellar
Library (http://www.astro.mat.uc.pt/BaSeL) compiled by Lejeuene et al. (\cite{lejeuene}).
We considered two luminosity classes (dwarfs and giants) assuming only solar
metallicity. The sensitivity of the results on the last parameter appeared to be
negligible.

As the first step, we investigated the dependence of the DCR phenomenon on the
ground-level temperature, pressure and relative humidity  for the values
 characteristic for our observatory. As could be expected, taking into
account previous results (e.g. Malyuto \& Meinel \cite{malyuto}), the
correlation between the effect and these parameters is negligible. Therefore, in
further computations we used
standard atmospheric temperature and pressure assuming 60\%
humidity, which is typical for our location in spring and early summer.
The first task was to check how the differential refraction depends on the
difference in the zenith distances of the binary components. It appeared that
for a 1$''$ difference for all spectral classes and all filters this
effect is of the order of 0.001$''$, if observations are made at 60$^{\circ}$
zenith distance

Taking into account the precision of our determination of the astrometric
parameters, we assumed further on that both components were at an
identical zenith distance.

\begin{figure}
\resizebox{\hsize}{!}{\includegraphics[bb = 50 175 515 630,clip]{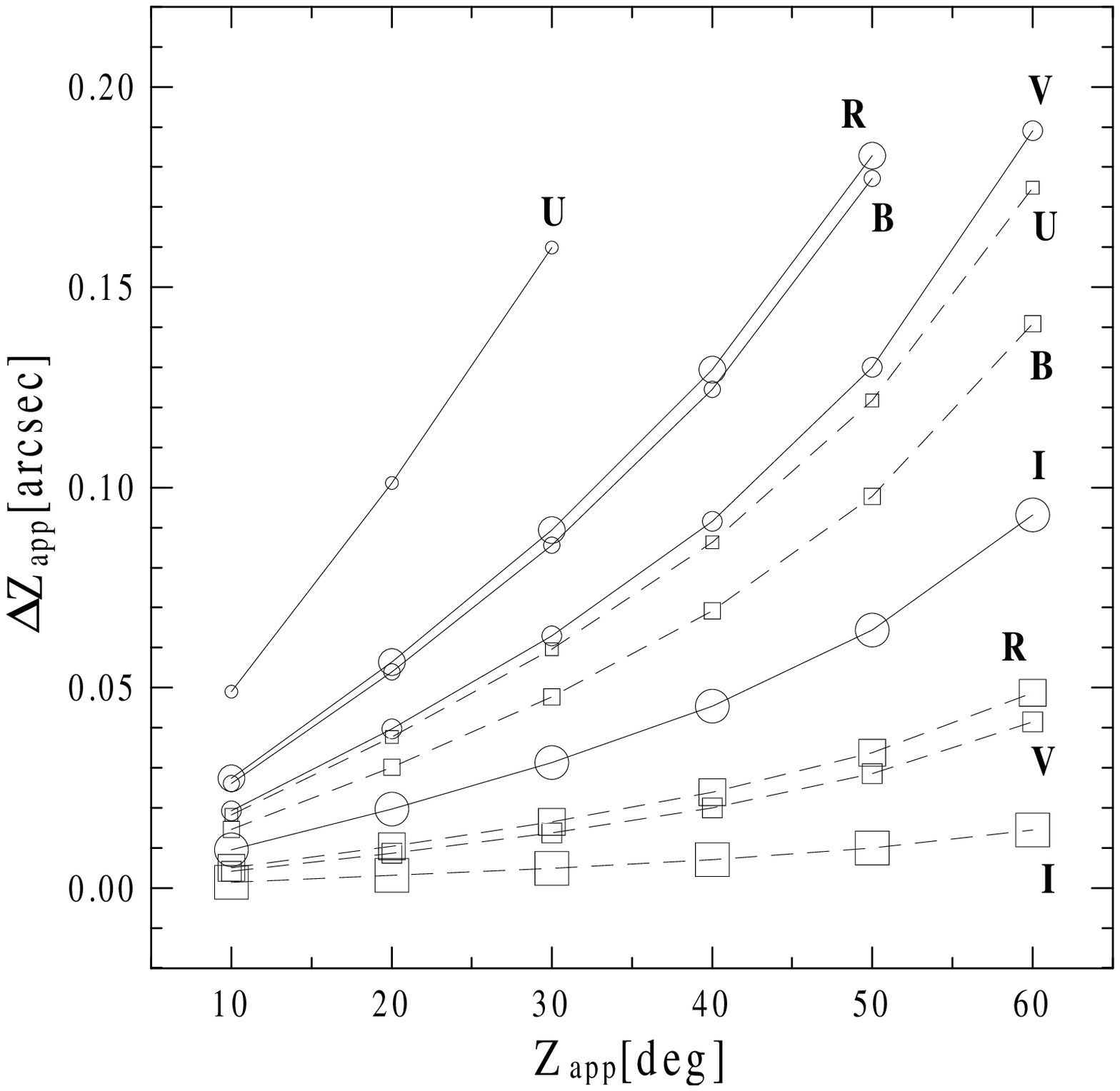}}
\caption{Atmospheric differential colour refraction versus apparent zenith distance of a pair of dwarf stars with a Solar value of Log(g). Open circles
connected by solid line: effect for binary star composed of O5 and M5
components. Open squares connected by dashed line: binary star with O5 and
F5 stars.}
\label{Fig3}
\end{figure}

Of the various results which we obtained, only the most important
are presented here. Fig.~\ref{Fig3} 
shows the dependence of the DCR effect on the zenith
distance of the binary star for two cases of the difference between spectral
types and for all filters which we were able to use. This result fully confirms
our decision to use only V,R,I filters in our observations. Although the O5-M5
difference between components makes it impossible to obtain reasonable
astrometric results, the interval of three classes gives an effect less than
0.02$''$ for a
zenith distance smaller than 40$^{\circ}$° for all three filters used. As could
be expected for the I filter the DCR effect is the smallest (less than 0.01$''$
for z$<$50$^{\circ}$°). The R filter gives the worst results, due to the fact,
that
the effective width of its spectral profile is about twice the width of the
profile of the V filter.

\begin{figure}
\resizebox{\hsize}{!}{\includegraphics[bb = 50 175 520 630,clip]{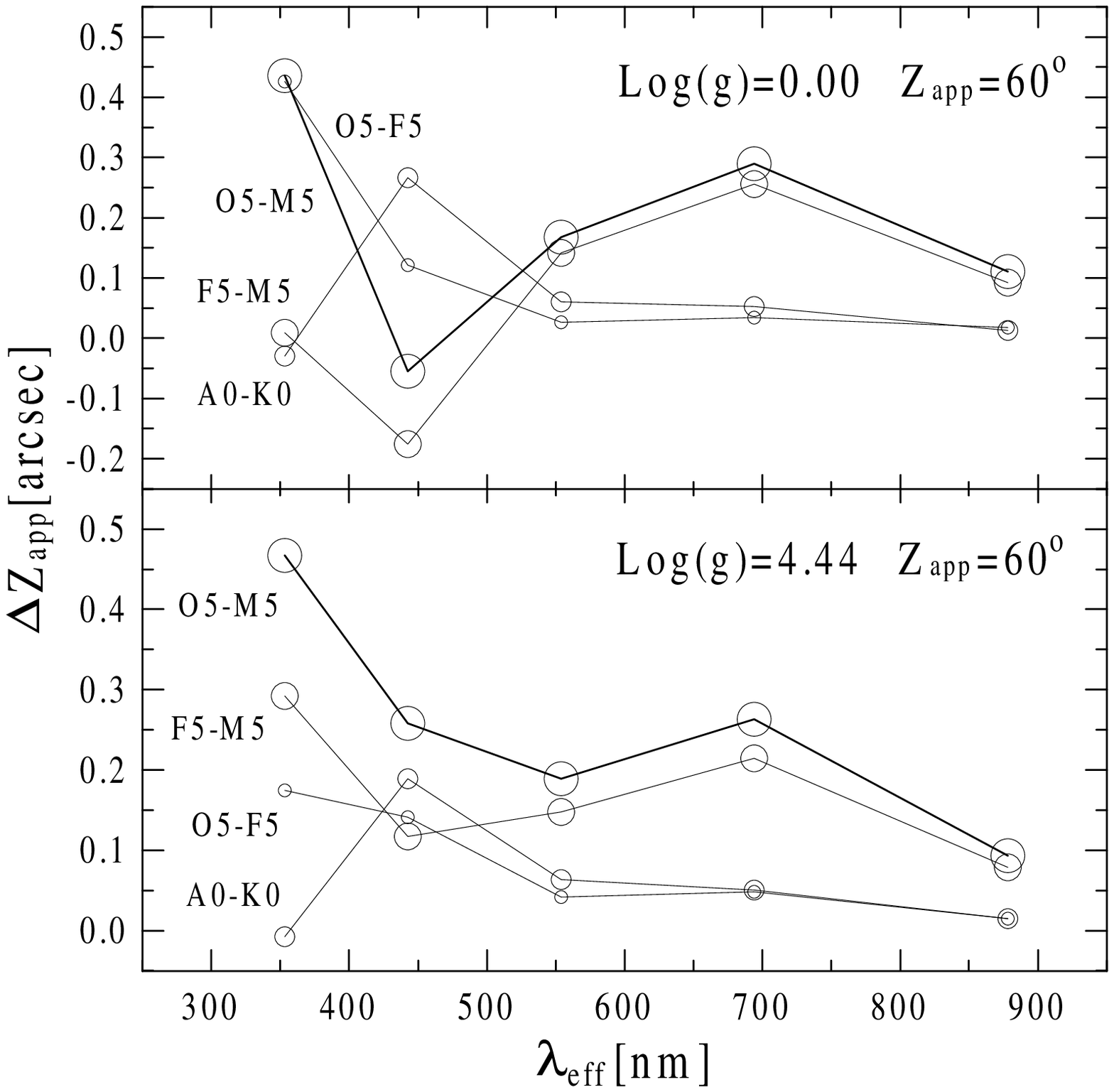}}
\caption{Atmospheric differential colour refraction for UBVRI filters for
an apparent zenith distance of 60$^{\circ}$ and two luminosity classes.
The X-axis gives the effective wavelength of the filters. Four different
cases of spectral classes of the components of the binary star are presented.}
\label{Fig4}
\end{figure}

Fig.~\ref{Fig4}
presents the DCR effect at 60$^{\circ}$ zenith distance versus the
effective
wavelength of the filter for two luminosity classes of the binary components
(a pair of dwarfs and a pair of giants) and for various differences of the
spectral types of the components. The results for the pair dwarf+giant is not
so different as to need to be presented. As can be seen, the sequence of the
filters
ordered according to increasing DCR effect (I,V,R) generally does not depend
on the spectral or luminosity class. A smaller effect
for these three filters can be expected for earlier spectral types. A late
spectrum implies a  more complex spectral energy distribution in the
long wavelength
range. The DCR effect for the U,B filters appears to depend on the spectral and
luminosity classes in quite a complicated way. This is once another argument against
observing through the broad-band short wavelength filters.

Surprisingly, cases exist for which the effect for the least suitable U
filter is smaller than for B filter and of the order of the effect for the V,R,I
filters. It occurs for moderate spectral classes where the Balmer jump makes
the effective spectral profile of the U filter when combined with stellar
spectrum relatively narrow.
\section{Results and discussion}
Table 1 presents the results of our speckle interferometry. It gives
separations,
position angles and magnitude differences between components for the
observation moments expressed in Besselian years. In cases when more than
one observation of a given binary was performed during a night the results were
averaged and the mean values were put under horizontal line. Errors of the
parameters were derived from LSQ fitting of a sinusoid to the mean power
spectrum of the speklegrams. They include the errors of determination of
the scale and position angle of our frames. It is not surprising that the
discrepancies of the parameters obtained from the LSQ approach are rather
underestimated due to the correlations between the parameters themselves. Only for
multiple observations can appropriate values of the errors be received. Errors
of the mean values of the parameters presented in Table 1 merge the two kinds of
cited above. To give a feeling of how high the real errors of the astrometric
parameters could be in cases when only a single observation of
the binary star was made, we compared errors the given by LSQ fitting with the
errors obtained from multiple observations. The correlations were quite good and
show that the real error of the separation should be about 3 times and the error
of the position angle about 4 times greater then the formal LSQ error.

Even superficial inspection of Table 1 shows that the astrometric results are
almost independent of filter, exposure time and PSF standard used, especially
taking into account the precision of our measurements. On the other hand,
magnitude
differences show serious discrepancies, particularly if the separation of the
components becomes comparable or smaller than our diffraction limit. This is
fully understandable in the case of LSQ fitting. Thus, only consistent multiple
results of the determination of the magnitude difference for the separation of a
pair larger than diffraction limit can give reliable information
(WDS04100+8042=ADS2963, WDS14139+2906=ADS9174, WDS18208+7120=ADS11311).
The remaining results should be treated only as rough approximations.

\begin{figure}
\resizebox{\hsize}{!}{\includegraphics[bb = 50 170 530 640,clip]{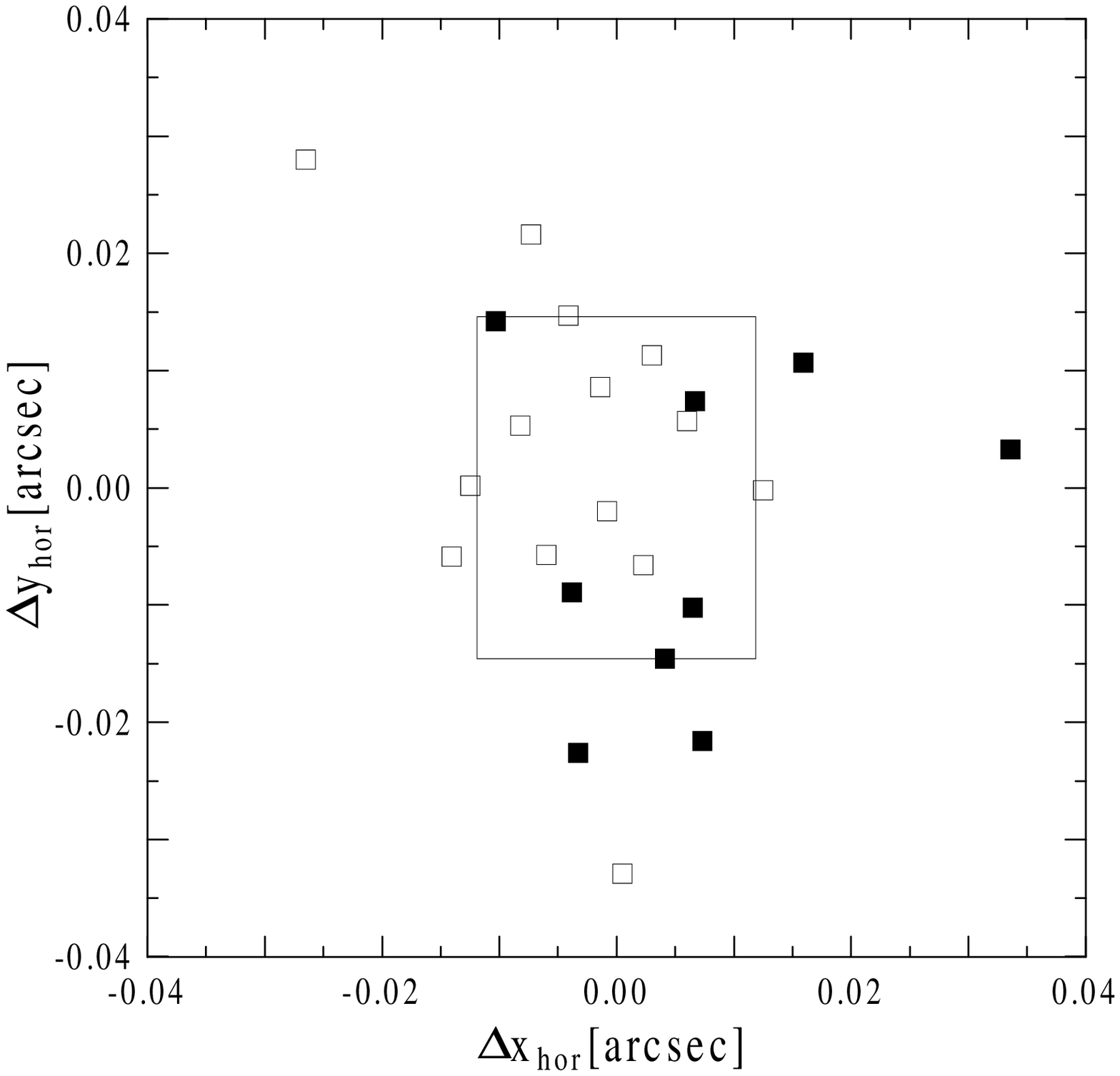}}
\caption{Dispersion of the observational points in the rectangular
horizontal-vertical coordinate system. The Y - axis is parallel to the vertical
coordinate and measures dispersion of the zenith distances. The figure presents
(O-C) differences between the mean (for at least two measurements) and actual
for a given star. The central rectangle shows 1$\sigma$ bands for x and y at
0.012 and 0.015 arcsec respectively. Open squares are for V,R filters and
filled symbols represent the I filter.}
\label{Fig5}
\end{figure}

To investigate statistically the possible biasing of our astrometric results by
the DCR effect, we analysed differences between the actual and the mean relative
positions of the components for multiple measurements. They were referred to a
horizontal-vertical reference system. The DCR effect should be manifested for
the vertical component of the dispersion since the consecutive measurements were
performed at different zenith distances. Fig.~\ref{Fig5}
shows these discrepancies
separated into two classes. For V,R filters (first group) the effect should
be on average greater than for the I filter (second group). As one can see, the
difference between vertical and horizontal dispersions is slightly significant
and any correlation between the magnitude of the effect and the filter could be
discern. The conclusion is that from the statistical point of view our
observational conditions and the differences between the spectral/luminosity
classes of the components were on average very favourable, producing negligible
DCR effect.

\begin{figure}
\resizebox{\hsize}{!}{\includegraphics[bb = 50 170 530 640,clip]{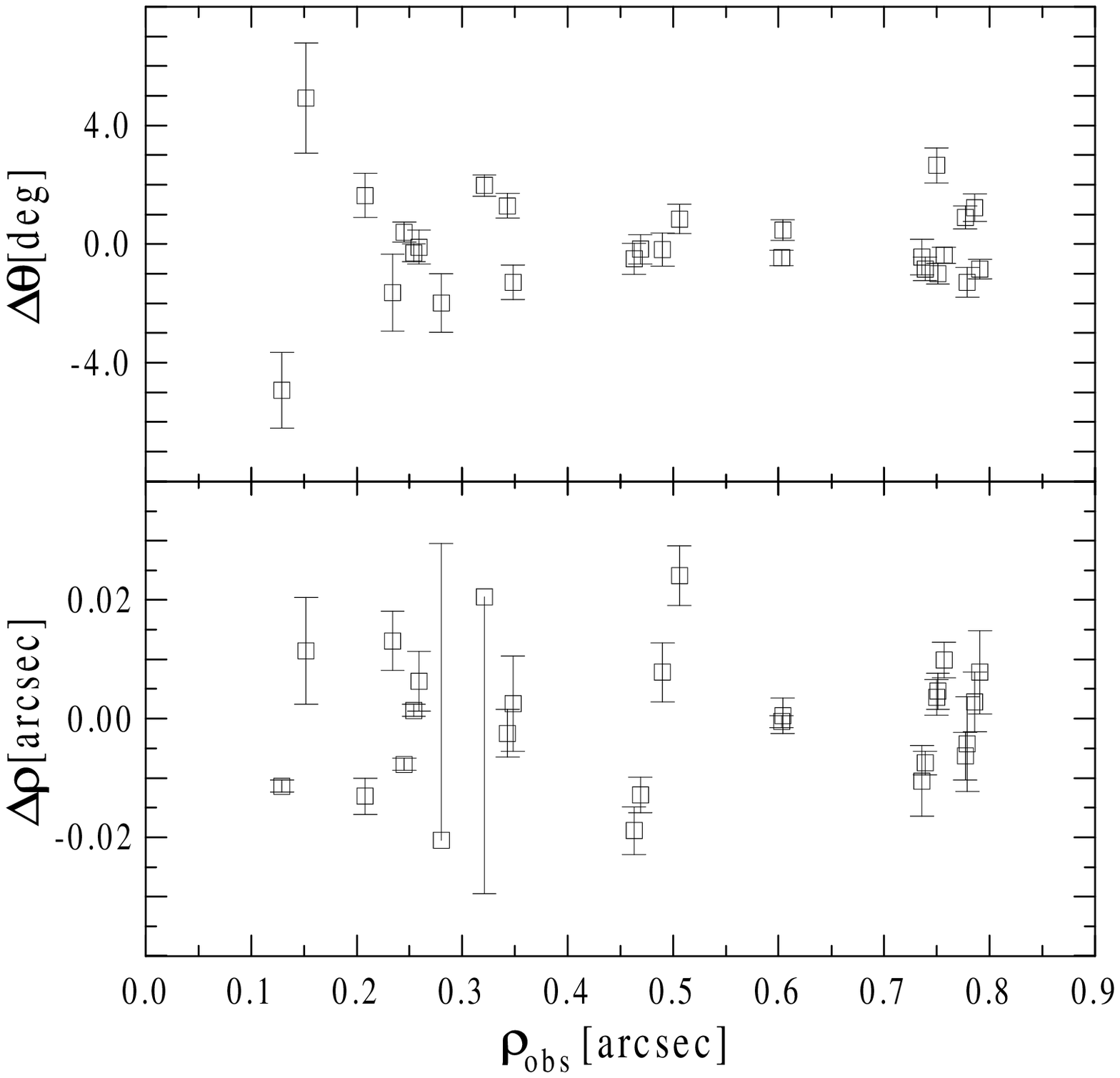}}
\caption{Dispersion of the position angles and angular separations as a function of the separation
between components. This figure presents $(O-C)_\rho$ and $(O-C)_\theta$ differences between the mean
(for at least two measurements) and actual values for a given star. The error bars were inferred from
the LSQ fitting procedure of the speckle power spectrum.}
\label{Fig6}
\end{figure}

Fig.~\ref{Fig6}
\begin{figure*}[!ht]
\centering
\includegraphics[width = 17cm, bb = 50 230 540 710,clip]{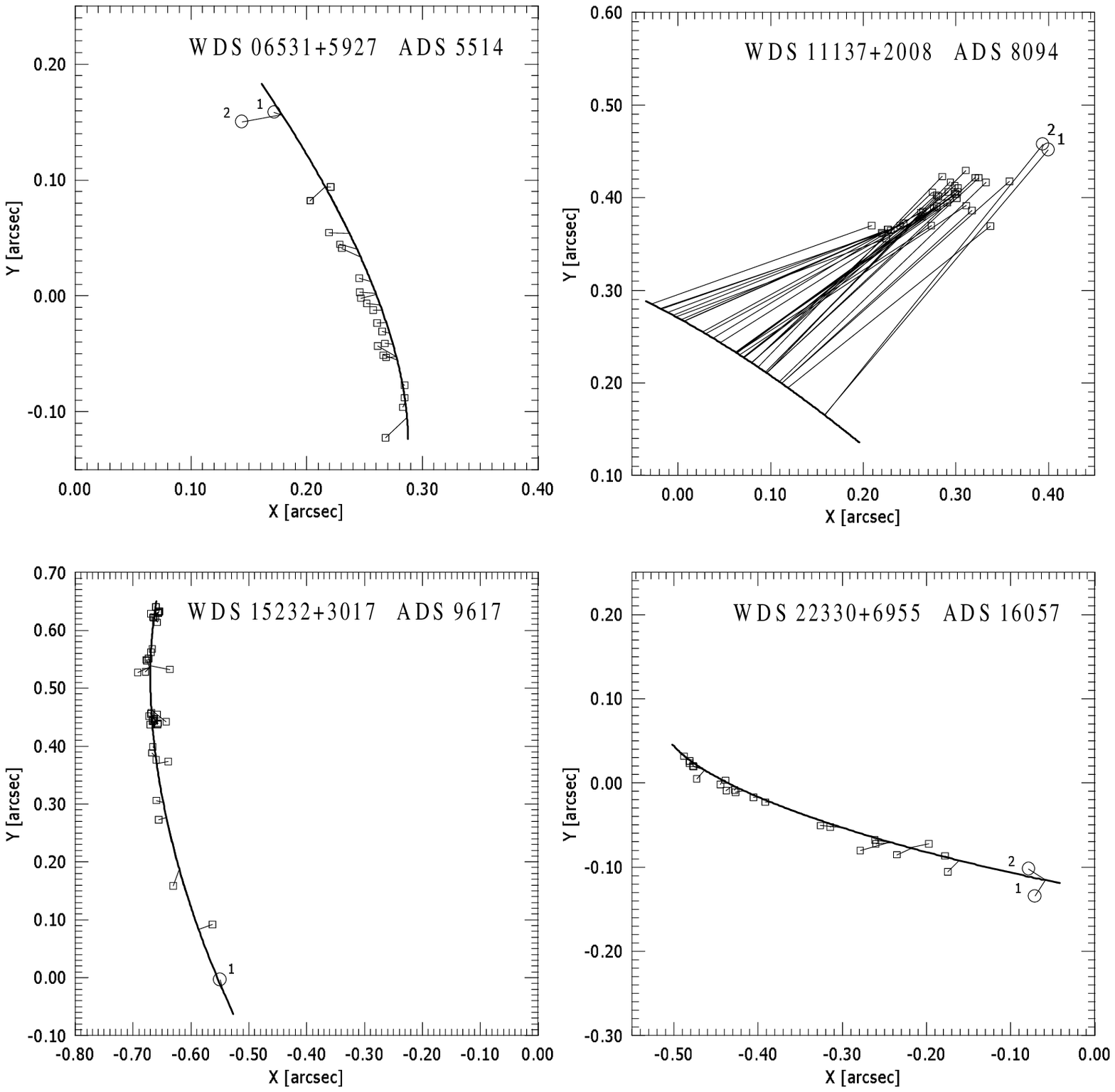}
\caption{Comparison of our astrometric results (open circles) with other speckle interferometric results
compiled in Hartkopf et al. (2004) (open squares) and actual orbits according
to the orbital elements listed in the 6-th Catalog of Orbits of Visual Binary
Stars (Hartkopf \& Mason 2003a). Line segments are drawn from the measured
positions to the ephemeris points. North is up and East is to the left.}
\label{Fig7}
\end{figure*}
 presents the uncertainties of $\rho$ and $\theta$ (defined as previously
but refereed to the polar reference system) versus the separation of the
components. The only trend is an increase of $\Delta \theta$ with decreasing
$\rho$, which is quite normal. The RMS values of the discrepancies are
0.012$''$ for $\rho$ and 1.8$^{\circ}\ $for $\theta$. If one wants to consider
results obtained for the
separations larger than, or at least close to, the diffraction limit one should reject
star WDS22330+6955=16057. Then, the mean dispersion of the separation is almost
unchanged, whereas the mean dispersion of $\theta$ drops to 1.2$^{\circ}$.
The precision of our measurements is comparable to typical values for 1 m class
telescopes.
Fors et al. (\cite{fors-b}) using a 1.5 m telescope presented an accuracy of
0.017$''$ and 1.5$^{\circ}$ respectively. Measurements by Scardia et al.
(\cite{scardia}) made with the PISCO
instrument (Prieur et al. \cite{prieur-a}) moved to Merate and attached
to 1 m telescope had an overall dispersion of about 0.01$''$.

To compare our position angles and separation measures with the
results of other observers and with ephemerides we took the
observational data compiled in the Fourth Catalog of Interferometric
Measurements of Binary Stars (Hartkopf et al. \cite{hartkopf-c}) and actual
orbital elements in the 6th Catalogue of Orbits of Visual Binary Stars
(Hartkopf \& Masnon \cite{hartkopf-a}). Fig.~\ref{Fig7}
presents such comparisons for 4
selected pairs having different
grades of orbit determination. An especially odd case is WDS1137+2008=ADS8094, for
which it seems that the "orbital" motion is a typical proper motion of two
separate stars.

\begin{figure}
\resizebox{\hsize}{!}{\includegraphics[bb = 50 170 530 635,clip]{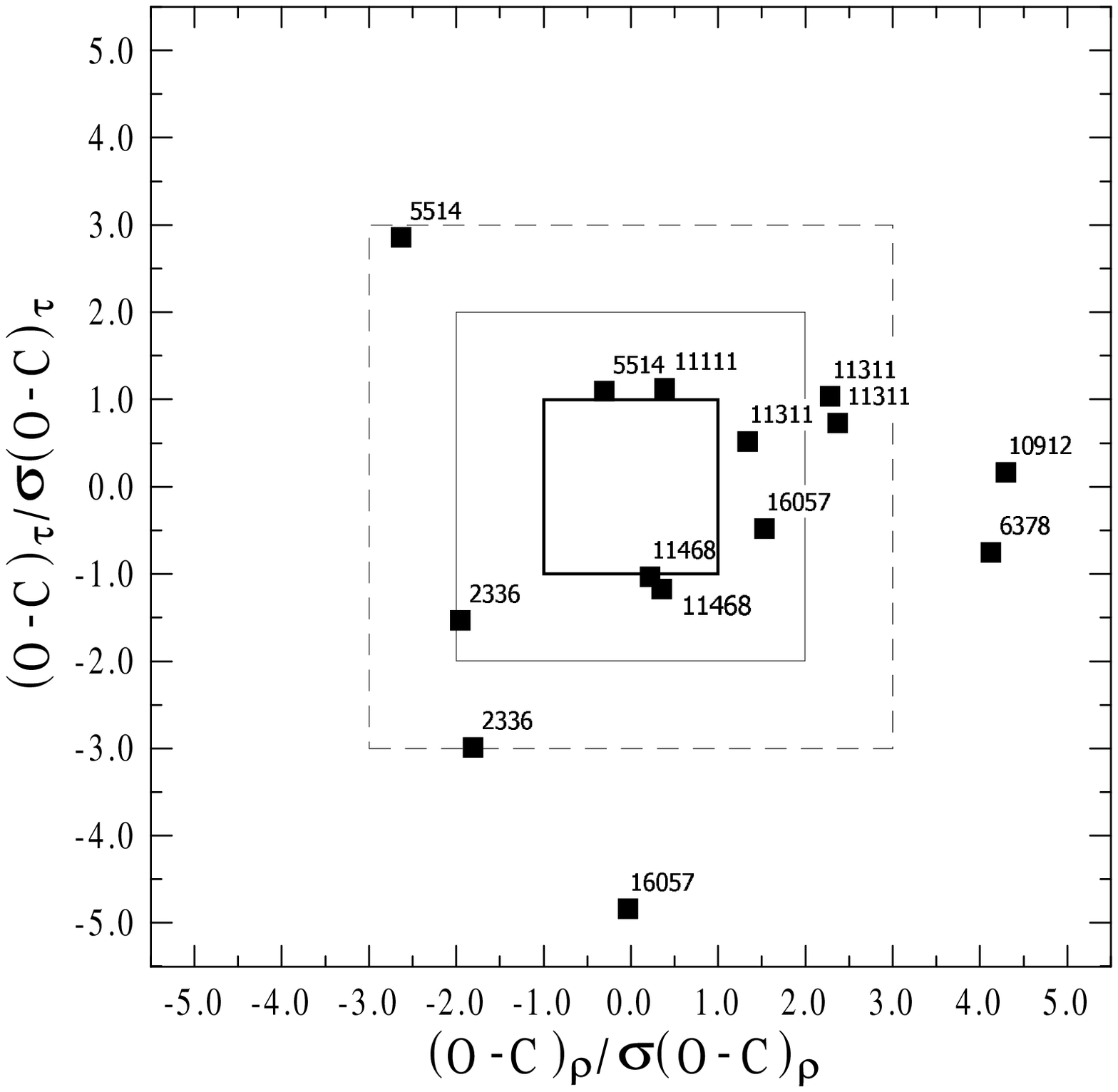}}
\caption{Relative (O-C) for a number of stars presented in the coordinate system
of angular separation $\rho$ and tangential displacement
$\tau$=$\rho$$(O-C)_\theta$. This plot shows ratios of the actual (O-C) values
to the RMS of (O-C) inferred from speckle results compiled in
Hartkopf et al. (2004) and actual orbits taken from the 6-th
Catalog of Orbits of Visual Binary Stars (Hartkopf \& Mason 2003a). Thick, thin
and dashed lined squares mark 1,2 and 3$\sigma$ boxes respectively. The
binaries are described by ADS numbers.}
\label{Fig8}
\end{figure}

To give general information of how good our results are, compared to
other results of speckle interferometry, we should not only analyse the
differences between the ephemeris positions and ours, but compare
these differences with the total behaviour of the O-C measure. A spectacular
example, which supports our approach, is WDS1137+2008=ADS8094 (mentioned above).
Thus, for our binary stars having orbits of grade better or equal to 4 and
which do not exhibit systematic temporal trends of the O-C measure, we computed our
differences between observed and ephemeris positions and divided them by the RMS
averaged O-C values for all the results cited in Hartkopf et al.
(\cite{hartkopf-c}). Fig.~\ref{Fig8}
presents our relative O-C discrepancies and 1,2,3 $\sigma$ boxes. Although
none of the points fit inside the 1 $\sigma$ square, only 3 measurements
of 14 are outside the 3 $\sigma$ box. Examining this result one should bear in
mind that a considerable fraction of the observations presented in Hartkopf et
al. (\cite{hartkopf-c}) were performed with a few meters class telescopes and
with correctors
of the DCR effect. As can bee seen from Fig.~\ref{Fig8} no substantial bias in the
angular separation or in the position angle exists in our data.

We also averaged direct O-C measures and computed their RMS values for the group
of binary stars described above but without WDS22330+6955=16057, for which the
angular separation is below the diffraction limit. For the position
angle we obtained a mean O-C value equal to -0.40$^{\circ}$ and the RMS of
2.89$^{\circ}$. In the
case of the separation, these values were 0.010$''$ and 0.046$''$, respectively.
Our O-C measures for a number of stars can be directly compared to the most recent
results received by Mason et al. (\cite{mason-a}, \cite{mason-b}) with 0.66 m
USNO refractor. On average our values of the discrepancies between observed
and ephemeris positions are close to the discrepancies given by Mason.

\begin{figure}
\resizebox{\hsize}{!}{\includegraphics[bb = 105 579 507 780,clip] {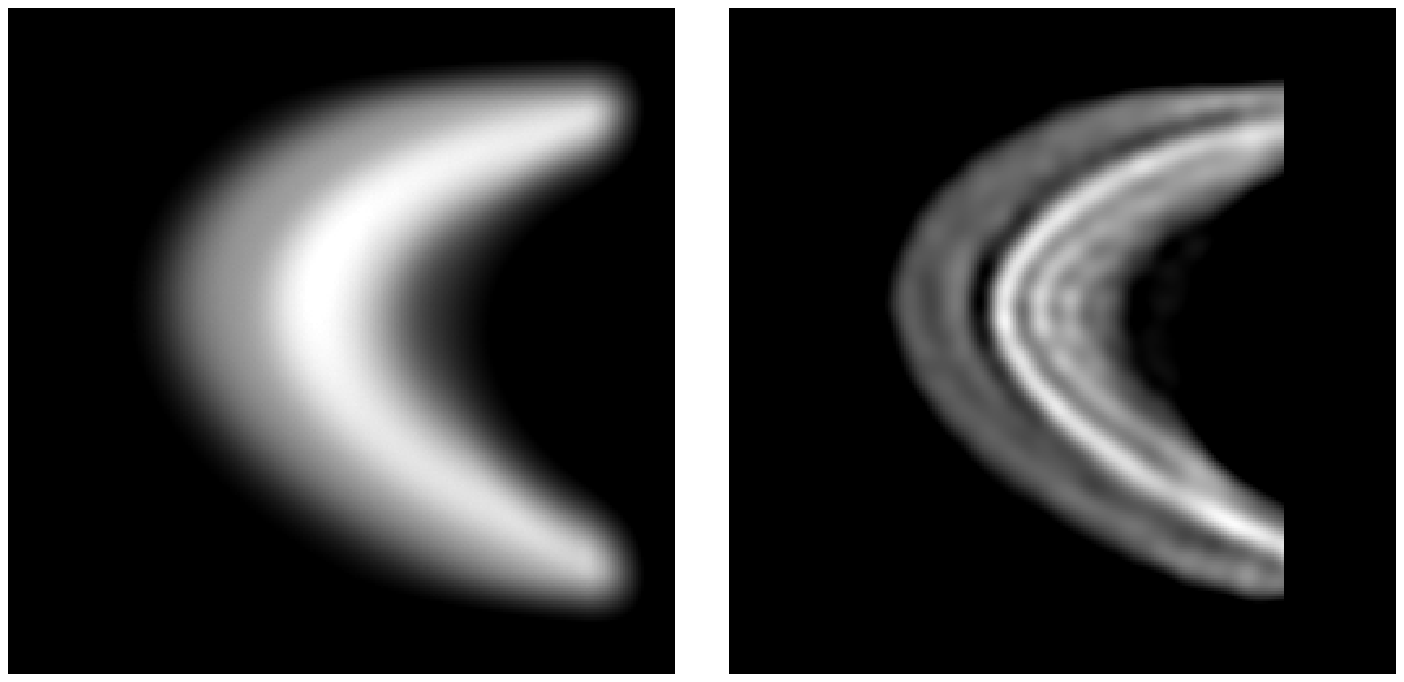}}
\caption{Example of using speckle interferometry with the cross-spectrum technique
to enhance  image resolution in the case of an extended object larger than the atmospheric
coherence patch. The figure presents a fragment of Saturn's rings with the disk of
the planet masked. Left panel: usual 15 sec exposure,
right panel: image reconstructed from 300 speckle images. The side of the frame has 20 arcsec.
North is up and East is to the left.}
\label{Fig9}
\end{figure}

As a by-product of our speckle interferometry we obtained the speckle
frames of bright extended objects, such as Saturn, Io, and Ganimede. Here we
present the results of speckle image reconstruction of a fragment of
Saturn's rings as an example of the enhancement of resolution degraded by seeing.
At the beginning we observed the entire ring structure using 512x512 frames
instead of 265x265 and attempted to reconstruct the image, but the result was
only slightly better than the usual image degraded by seeing. Therefore, we omit
this discussion here. During the night of 27 March 2003, we obtained 300
specklegrams of a half of the full ring with 0.05 sec nominal exposure time,
using the I filter and 10x microscope objective. As a PSF standard we used BS1910.
The problem of crucial importance during the data reduction was to
properly apodize the speckle frames. First, we prepared the mask, which removed
half of the planetary disk and apodized the borders of the image with a cosine
function.
Before masking, all the frames were shifted to overlap one another using the
cross-correlation method. After computing the mean power spectrum of the
apodized specklegrams we proceeded in the same way as for binary stars. Fig.~\ref{Fig9}
shows the reconstructed image (right panel) compared with the usual exposition
obtained during the same night for 2$''$ seeing (left panel). The size of the
frame is 20 arcsec, which is at least a few times greater than the average radius
of the atmospheric coherence patch. In this case a partial reduction of the
seeing effect was obtained but the image resolution was substantially enhanced.
An additional effect is a distinct, though not strong excessive deconvolution
of the ring structure, manifesting itself by too high contrast between the
rings and divisions and knotty artefacts visible in the image. Nevertheless, the
bright A,B rings as well as Encke and Cassini divisions
are quite clearly visible.
\section{Conclusions}
We have presented the results of the first speckle interferometry
 observations of visual binary stars made with the 50 cm
 Cassegrain telescope of the Jagiellonian University in Cracow.
 By adding a simple image projector with a
 microscope objective to the standard solution for the broad-band CCD photometry,
 we obtained good spatial sampling of the frame. Using standard V,R,I filters
 without a DCR effect reducer forced us to perform a theoretical
 analysis of the influence of this phenomenon on the astrometric parameters.
 Statistical analysis of our observational data confirmed the model results.
 For V,R,I filters the DCR effect for moderate zenith distances was almost
 negligible, compared to the precision of our astrometry, which was about
 0.014$''$ taking into account both separation and position angle measures.
 Such a value of the dispersion is rather typical for 1 m class telescopes
 if one uses LSQ fitting of the mean power spectrum by a sinusoid function.
In evaluating our precision, one should bear in mind that we used
a limited number of speckle images (typically 300 in one trial) and
  exposure times longer than 0.04 sec.

Our approach to the problem of the reconstruction of the phase information based
on iterative LSQ analysis of the 8 sub-planes of the cross-spectrum gave
good results, not only resolving the quadrant ambiguity but allowing us to
retrieve almost diffraction limited images of the extended objects. As an
example an image of a fragment of  Saturn's rings was presented.

The analysis of the O-C measures of our data shows general agreement with the
results of other projects observing binaries by speckle
interferometry. Although our telescope is one of the smallest used for such a
task, our results are hardly inferior to the results obtained with better
equipped 1 m class telescopes. Thus, our ability to perform reasonable
astrometry are quite promising. Unfortunately, the precision of our relative
photometry of binary components were rather poor (the order of a few tens of
magnitude) and requires some improvements in the data reduction.

From our results for WDS1137+2008=ADS8094, which has orbital parameters of
the worst grade (Hartkopf \& Mason \cite{hartkopf-a}), it is rather obvious that the relative
trajectory observed until now contradicts the presumed double nature of
this object.
\begin{table*}
\caption{Results of speckle interferometry for binary stars.
Consecutive columns give: WDS and ADS numbers; fraction of the Besselian Year;
nominal exposure time in msec, filter used, BS or SAO (in italics) number of the
PSF standard; angular separation in arcsec together with the error and (O-C) value;
position angle in deg measured from North towards East for the 2000.0
epoch, error and (O-C) value for the position angle, grade of the orbital elements (G) according
to the 6-th Catalog of Orbits of Visual Binary Stars (Hartkopf \& Mason 2003a), magnitude difference
between components ($\Delta$) together with its error and the number (N) of the note concerning the observational
conditions.}
\label{Table}
\includegraphics[width = 1.0\textwidth, height = 0.9\textheight, bb = 50 50 540 790,clip]{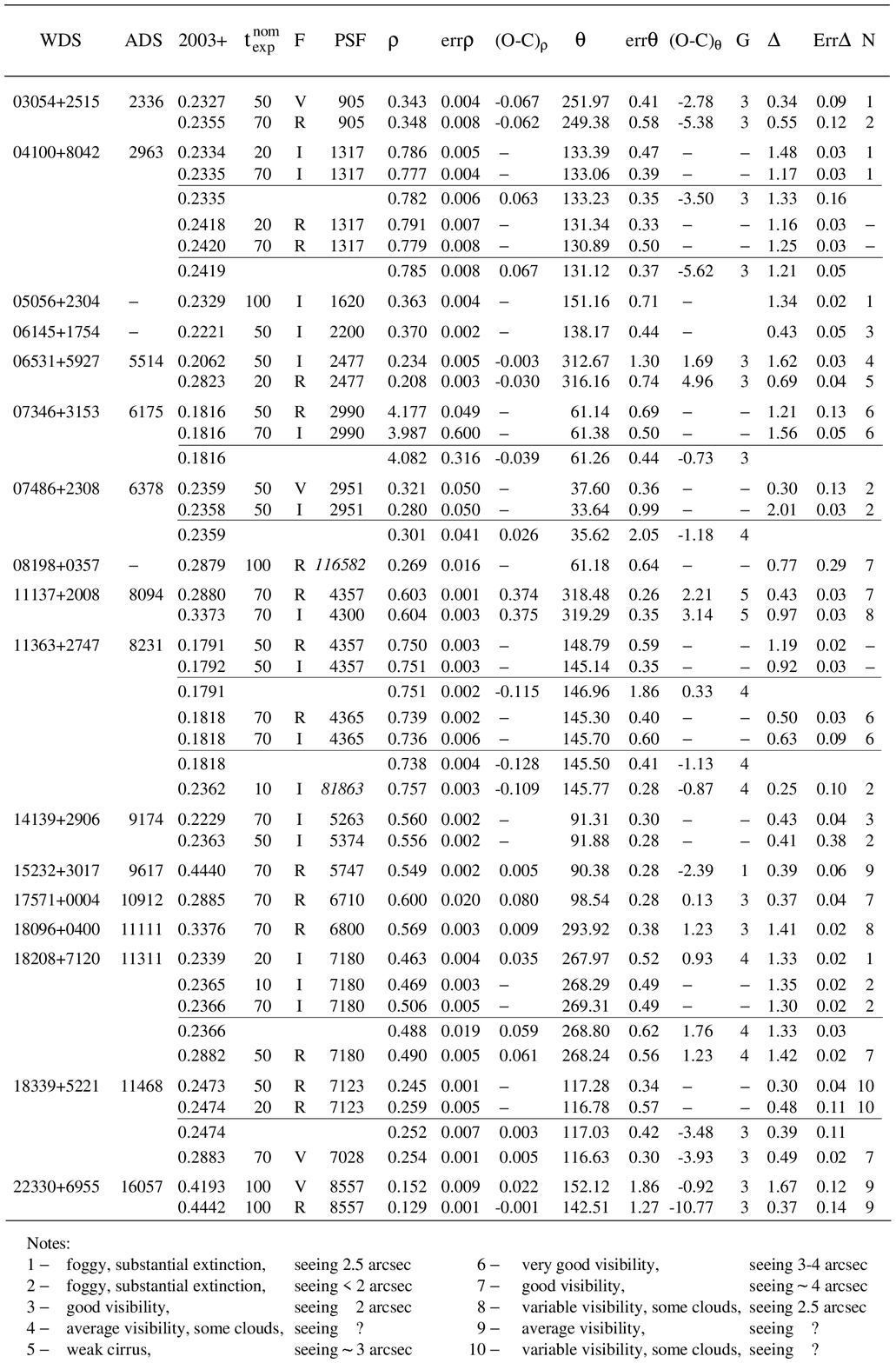}
\end{table*}

\begin{acknowledgements}
We would like to thank M. Siwak for kindly supplying extinction coefficients
for Cracow in the U,B,V,R,I filters.
\end{acknowledgements}

\end{document}